# A 2D self-organized percolation model for capillary impregnation

A. K. Nguyen[1,2], C. B. Trang[1,2], E. Blond[1], E. De Bilbao[3], T. Sayet[1] and A. Batakis[2]

[1] *Univ. Orléans, LaMé (EA7494, Univ. Orléans, Univ. Tours, INSA CVL), France, {anh-khoa.nguyen, eric.blond, thomas.sayet, cong-bang.trang}@univ-orleans.fr*
[2] *Univ. Orléans, Insitut Denis Poisson (UMR CNRS, Univ. Orléans, Univ Tours), France, athanasios.batakis@univ-orleans.fr*
[3] *Univ. Orléans, CEMHTI (UPR CNRS), France, emmanuel.debilbao@univ-orleans.fr*

**Abstract** —A two-dimension extension of the **S**elf-organized **G**radient **P**ercolation (**SGP**) method initially developed for the one-dimensional simulation is proposed. The initialization in the two directions is considered as the analytic solution of the 2D (homogeneous) diffusion equation. The evolution of the saturation front is assumed to be the evolution of both standard deviations in each direction. The validation of the implementation is done by comparisons between SGP and finite element results.

**Keywords** — Impregnation, porous media, gradient percolation, simulation.

## 1. Introduction

To simulate the unsaturated (non-reactive) impregnation in a porous medium, classical numerical methods (FEM, theta-method, etc.) have often been employed to solve the Richards' equation [1], [2]. Yet, the CPU time and spurious oscillations become prohibitive, as for instance, when the impregnation is coupled with chemical reactions. That is why a drastically different, and more reliable approach, called **S**elf-organized **G**radient **P**ercolation (**SGP**) method, was proposed [3].

The initialization of the algorithm is driven by an analytic solution of the (homogeneous) diffusion equation. It is a convolution between a **P**robability **D**ensity **F**unction (**PDF**) and a smoothing function [4]. The evolution of the capillary pressure profiles with time is reproduced by the self-evolution of the standard deviation of the PDF. To test this model, the capillary pressure profiles and the mass gain curve have been confronted with those obtained by FEM and experimental measurement, respectively [3].

The main goal of the present work is to extend the SGP method to the 2D unsaturated impregnation only considering the non-reactive case. The two-dimensional SGP algorithm has the same structure than the one-dimensional one. On the mathematical side, the analytic form is a convolution between a PDF of two directions and a smoothing function. In addition, the evolution of the saturation front is proposed to be the evolution of the two standard deviations of the PDF in each direction. The visualizations and comparisons between the numerical results from SGP method and FEM are given.

## 2. Self-organized Gradient Percolation (SGP) method for impregnation

The gradient percolation method is a probabilistic approach to predict a spatial evolution. The idea for using this method here is to avoid the numerical difficulties of the resolution of the Richards' equation derived from the mass balance combined with Darcy's law [3]. Richards' equation is a nonlinear **P**artial **D**ifferential **E**quation (**PDE**), which requires a small-time increment and an appropriate fine space discretization to make sure the accuracy of the results. However, a small-time increment leads to a high computational cost and, moreover, a very small time-increment can give rise to spurious oscillations that impact the accuracy of the results [5]. The main goal for the development of this method is to reduce the CPU time and the number of parameters actually used for classical models. The **S**elf-organized **G**radient **P**ercolation (**SGP**) method, which is based on the gradient percolation method, has been developed to reach these goals.

Two fundamental types of percolation are broadly used to simulate physical phenomena, the bond percolation and the site percolation [6]. The site percolation can be applied to the simulation of the non-reactive impregnation process for the determination of the saturation, first considered as a state variable describing the state ("occupied" or "empty") of each local pore of the media.

Considering a random network defined by numerous cells/squares, function $X(z)$, where $z$ is the position of the cell, is a random value uniformly distributed on the interval [0,1]. A site declared to be "occupied" or to be "empty" by the liquid is defined in comparison with constant probability $p$ as:

- site $z$ is "occupied" when $X(z) \geq p$;
- site $z$ is "empty" otherwise.

In classical percolation model, probability $p$, which indicates the state of a site, is independent of the site, it is the same value for the whole domain. However, physics implies that the state of a site is given by capillary pressure, which is not constant on the lattice.

From physics point of view, the **C**apillary **P**ressure **C**urve (**CPC**) is at equilibrium only under the gravity field. The capillary pressure is the equilibrium pressure, which is determined by the difference between the non-wetting and the wetting phases. In the capillary rising test, the saturation profile along the vertical direction (Figure 1), considered as a function of the level (height) with respect to the liquid regarding Jurin's law, is an image of the CPC at each time $t$. That is why we propose to transform probability $p$ into a function of the location of the site. This can be done by adapting gradient percolation theory developed by Pierre Nolin [7]. Choosing $p$ as a function of the location allows reproducing the impregnation gradient, but the results will not respect the continuity of the liquid phase. Moreover, it is impossible to manage the boundary conditions. It is thus proposed to define a convolution product to satisfy the continuity and the possibility to take into account various boundary conditions. Finally, the local saturation is obtained through two steps: (1) local state $X$ is determined by the classical percolation and (2) employing the convolution product gives:

$$S(z) = X(z) * \delta(z) \tag{1}$$

where $\langle * \rangle$ is the convolution operator and $\delta$ is a smoothing function (spatial weighted average).

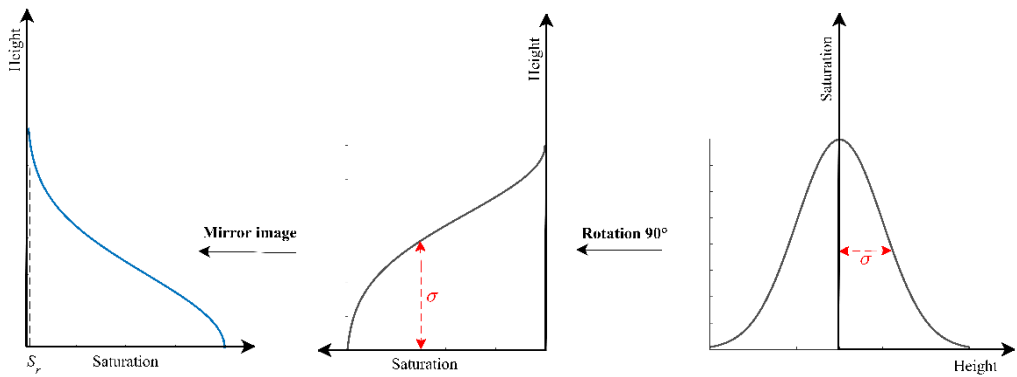

Figure 1 – The saturation profile as a probability density function

The saturation profile at initial time $t_0$ is then proposed by the following expression [4]:

$$X(P_{cap}, t_0) = S_r + (S_s - S_r)exp\left(-\frac{|P_{cap} - P_{cap,t_0}^{S_{max}}|^m}{m\sigma^m}\right) \qquad (2)$$

where $S_r$ and $S_s$ mean the residual and maximum saturation, respectively, $P_{cap}$ is the local capillary pressure, $P_{cap,t_0}^{S_{max}}$ is the minimum pore pressure to initiate the impregnation, $\sigma$ is the standard deviation of the PDF, $m$ is the empirical parameter to switch between probability distributions to fit the CPC (e.g. $m = 1$ and $m = 2$ designate Laplace and Gaussian distributions, respectively). From the physics point of view, $m$ is a dimensionless parameter which depends on the porosity network (such as porosity, tortuosity, etc.).

To reproduce the physics of impregnation, it is necessary that the model can evolve autonomously with time following the physical laws of the problem. In order to ensure this self-evolution along time, a relationship between the standard deviation and the capillary pressure is proposed [3]:

$$\sigma_n(P_{cap}) = \sigma_{n-1}(P_{cap}) + A\left(\frac{P_{cap,t_0}^S}{P_{h,n}} - 1\right) \qquad (3)$$

where $n$ designates the time increment and $A$ designates a kinetic constant ($m \cdot s^{-1}$).

Finally, there are only three parameters required for the SGP algorithm: $A$ ($m \cdot s^{-1}$), $P_{cap,t_0}^S$ (Pa) and $m$ (dimensionless). To identify these three parameters, the inverse identification method [3] is employed by fitting the mass gain curve obtained from the SGP method with the experimental one and then is validated by comparison with FEM simulation if the capillary pressure profiles are known. Then, $A$ is closed to the intrinsic permeability (but not fully explained today) while $m$ is directly linked to the shape of the capillary pressure curve and can be taken as the inverse of the parameter of van Genuchten's model. Therefore, only unknown $P_{cap,t_0}^S$ is the minimum pore pressure to initiate impregnation.

## 3. SGP method for 2D problem

Since the SGP algorithm presented above for the (quasi) one-dimensional case permits to reduce the CPU time and to avoid the spurious oscillations [3]. Its extension to the 2D case thus keeps working on and is straightforward to construct: the main parameter will no longer be the height but the total pressure. It is to note that the Richards' equation, where the height parameter is replaced by the total pressure parameter since from the physics point of view in the general case, is written as [3]:

$$\phi\frac{\partial S}{\partial t} = div\left(\frac{K(S)}{\eta}\underline{grad}(P_{tot})\right) \qquad (4)$$

where $\phi$ is the porosity (dimensionless), $S$ is the saturation (dimensionless), $K(S)$ is the nonlinear function of saturation ($m^2$), $\eta$ is the dynamic viscosity ($Pa \cdot s$), $\rho_W$ is the mass density of the wetting liquid ($kg \cdot m^{-3}$); $\underline{grad}(P_{tot}) = \underline{grad}(P_{cap} + \rho_W gz)$ is not a spatial description, but rather depends on the direction of gravity, where $P_{tot}$ is the total pressure ($Pa$), $g$ is acceleration due to the gravity ($m \cdot s^{-2}$) and $z$ is the height along the gravity direction ($m$).

### 3.1. Stepwise modelling procedure

Three main steps of the 2D SGP algorithm are presented as follows.

**Step 1:** Based on the main idea of solving the 1D (homogeneous) diffusion equation [3], the analytic solution of the 2D case, even in the general case, is pointed out to be a convolution between a Probability Density Function and a smoothing function. To be physically meaningful, the initialization is thus taken to be the analytic solution of the (homogeneous) diffusion equation (only considering initial time $t_0$) in the vertical and horizontal directions $(z_1, z_2)$ as the following:

$$X(P_{tot}(z_1), P_{tot}(z_2), t_0) = S_r + (S_s - S_r) exp\left(-\frac{|P_{tot}(z_1) - P_{tot,t_0}^{S_{max}}|^m}{m\sigma(z_1)^m} - \frac{|P_{tot}(z_2) - P_{tot,t_0}^{S_{max}}|^m}{m\sigma(z_2)^m}\right) \quad (5)$$

where $P_{tot}, P_{tot,t_0}^{S_{max}}$ mean the total pressure and the minimum pore pressure to start the impregnation assuming to be the same in each direction, respectively, $m$ is the empirical parameters defining the shape of the distribution in each direction and $\sigma(z_i)$ is the standard deviation function along the $i$-direction.

To take into account Eq. (1) and Eq. (5), we employ the dedicate function "norminv" in Matlab and then a multistate gradient percolation model is then given.

**Step 2:** In the next time steps, we assumed that the evolution of the capillary front along each direction is the evolution of each standard deviation, respectively. In particular, standard deviation $\sigma^n(z_i)$ along $i$-direction at timestep $n$ can be written as:

$$\sigma^n(z_i) = \sigma^{n-1}(z_i) + \alpha(P_{tot}^n(z_i)) \quad (6)$$

where $\alpha(P_{tot}^n(z_i))$ means the evolution of the standard deviation at timestep $n$ along $i$-direction. The link with the Poiseuille's equation has been pointed out in the 1D capillary rising case, which gives:

$$\alpha(P_{tot}^n(z_i)) = \frac{\gamma D^2}{32\eta} P_{tot}^n(z_i) \quad (7)$$

where $A = \gamma D^2/32\eta$ depends on capillary diameter $D$ $(m)$ and $\gamma$ is the surface tension $(N/m)$.

**Step 3:** In the final step, note that the motion of liquid here is due to driving force $\underline{grad}(P_{cap} + \rho_W gz)$. Obviously, only in the capillary rising case, the simulation will stop when there is no difference between capillary pressure and hydrostatic pressure [3]. Nevertheless, in the general case, the simulation will not be able to stop until there is no more liquid on the boundary of the source. For example, the horizontal impregnation is never stopped, since there is no effect of the gravity in this direction. In the numerical practice, we thus suggest comparing the difference between the mass of liquid in the container (denoted Π) and the mass gain in time in the porous medium (denoted $m_t$). It leads to that the SGP algorithm will stop when there is no difference in this comparison.

### 3.2. Boundary conditions

$\partial\Omega_i$ corresponds to the three types of boundary conditions [3]. At the initial timestep, depending on the treated problem, it is considered that almost one surface of the porous sample is in contact with the surface of liquid. This leads to impose that concerning row (first row) of the matrix is fully filled with liquid. A specific boundary condition will be applied on each interface. The choice of the boundary conditions depends on the interpretation of the physical phenomena.

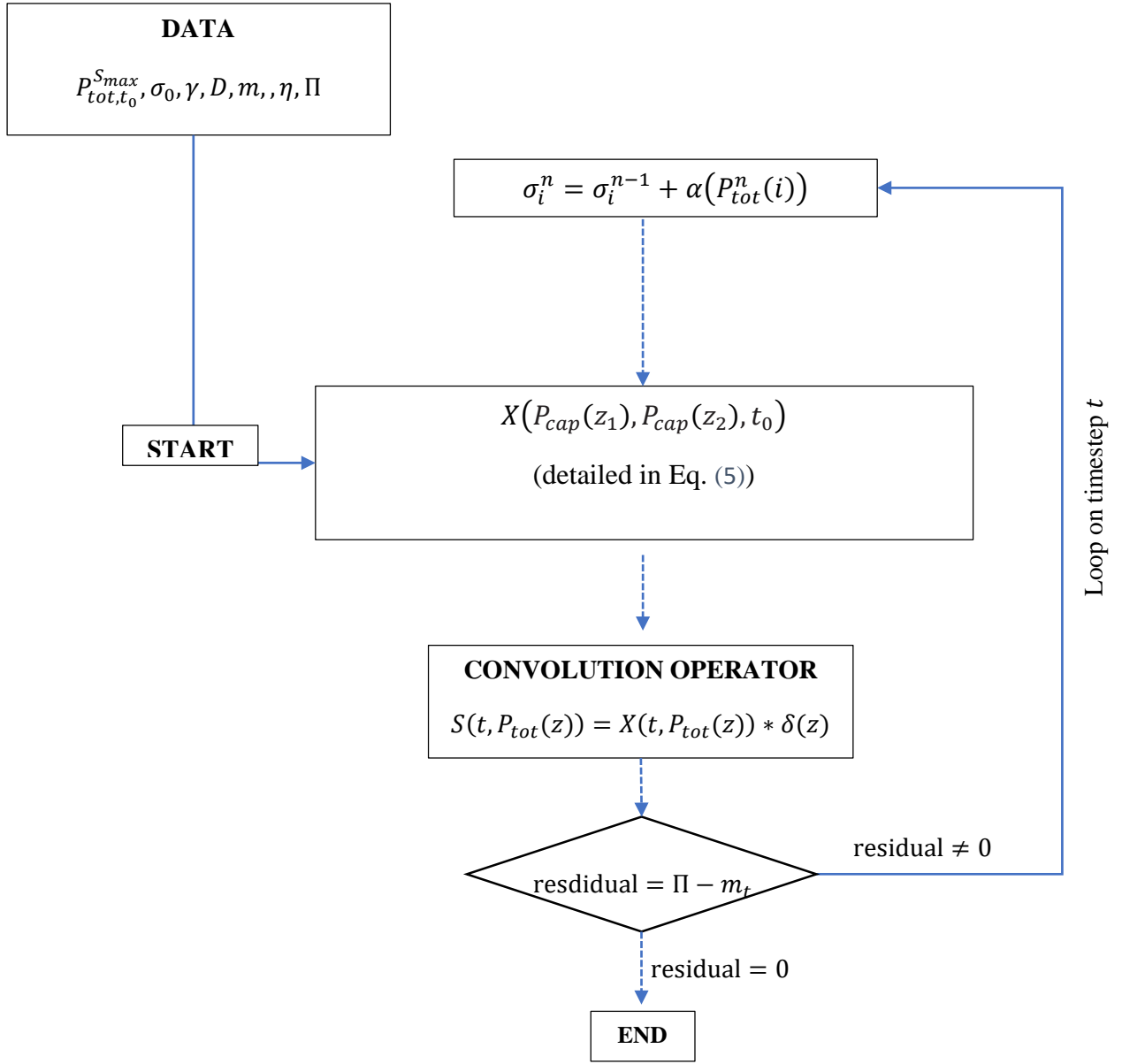

Figure 2 – Flowchart of the SGP algorithm for the 2D problem.

## 4. Application

In order to numerically simulate the test using FEM and the SGP algorithm, we decided to build the two models implemented both in (2D) Abaqus standard (v. 6.14) and Matlab (v. R2019a) software, respectively. To compare the CPU time of the two methods, it is essential that both models have to be built with the same mesh size, time increment, dimensions and boundary conditions.

### 4.1. Description of the test

For the application of the SGP algorithm extended to the 2D simulation of the impregnation, considering a numerical example, a cylindrical porous sample is in (partial) contact with a liquid at the bottom surface. The height and width of the sample are 0.04 m and 0.035 m (in Figure 3) respectively. The porous material is Alumina 99% and the impregnated liquid is glycerine.

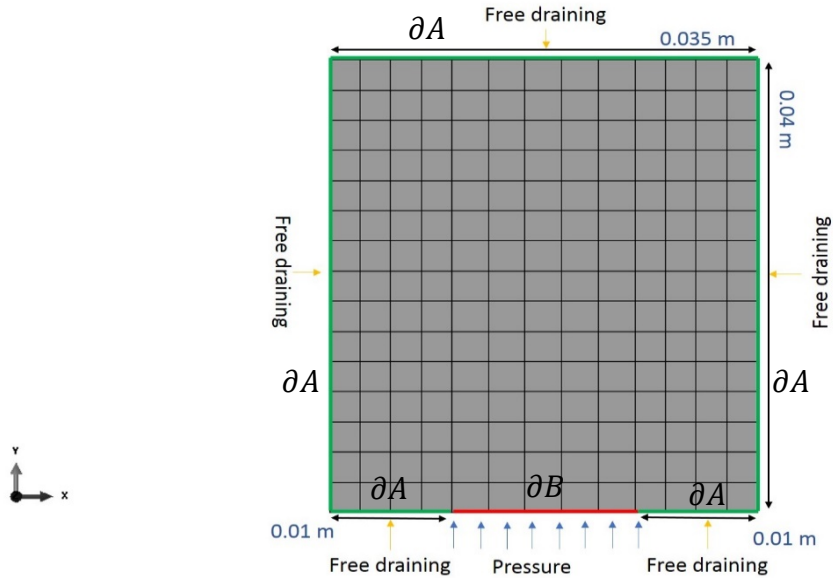

Figure 3 – The specific dimensions and boundary conditions of the numerical test are used both for the models using FEM and SGP method. The mesh size is 0.03 m (along surface $\partial B$), 0.025 m and 0.0267 m (along surfaces $\partial A$ in the horizontal and vertical directions, respectively).

Due to the lack of experiment data, the validation of the performance of the 2D SGP algorithm is made by comparisons of the local saturation of each square of the model using FEM with the one of each square with the same coordinates of the SGP model. Obviously, the number of the squares of each model is equal to the number of the squares in row multiplied with the one in the column. It leads to compare a very large number of squares in the case of very smooth mesh. To be simple in our ongoing study, we decided to consider the model with a coarse grid, but, of course, must respect the condition between time step and mesh size to avoid the spurious problems [5].

For the boundary conditions in both models, there is an imposed pressure on boundary surface $\partial B$ and the free draining on boundary surfaces $\partial A$ used in the simulation (Figure 3). In particular, for the first one, the liquid is in contact with surface $\partial B$, so it corresponds to an imposed total pressure; for the second one, to make sure that the liquid can be free to flow out of these surfaces, the drainage-only boundary condition is imposed on surface $\partial A$. The "expression" of these boundary conditions in the SGP model is done thanks to the convolution procedure [3].

### 4.2. Results and comparison

Table 1- CPU time and time ratios between FEM and SGP method.

| Durations (in seconds) | CPU time (in seconds) | | Time ratios |
|---|---|---|---|
| | FEM | SGP method | |
| 142 | 15 | 0.075 | 200 |
| 4254 | 48.3 | 0.5367 | 90 |

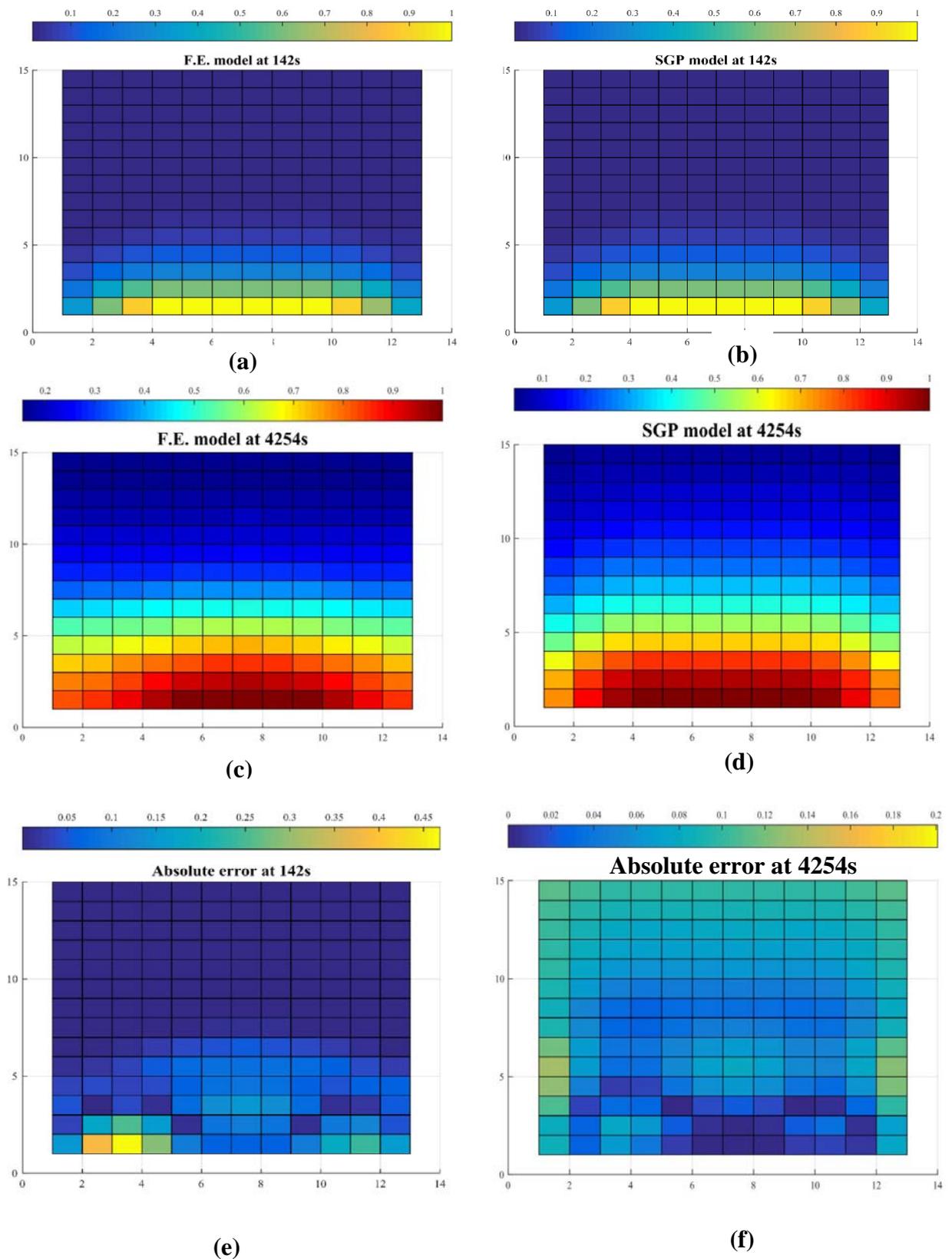

Figure 4 – Simulations using FEM and SGP method for the evolution of the saturation fronts at different time from (a) to (d). The absolute errors (e) and (f), calculated by the absolute minus of the local saturation at a square

of the model using FEM with the local saturation of the square with the same coordinates of the SGP model at different time, is resulted in the scale color bar.

The computational costs for the test for the SGP method and FEM are obtained by running the software on 2.60 GHz, x64-based PC with 32 GB of RAM. Obviously, the computational time of the SGP algorithm is significantly lower than the one of FEM (Table 1). This variation in the ratio is due to the difference in the evolution of time step between both methods [3].

## 5. Conclusion

In this work, our ongoing study of the extending 2D SGP method to the impregnation process is proposed. The preliminary visualizations from the FEM and SGP method (Figure 4 (a) to (d)) seems to indicate that the behaviour of the evolution of the saturation fronts from the two methods are the same at the first time steps (Figure 4 (e) and (f)).

Hence, there is still a huge work to do in order to study the mesh sensitivity, the impact of the time step, the impact of each of the parameters required for the SGP model and their link with the physics (properties of the material). For the perspectives, the SGP algorithm is expected to adapt itself to taking into account the kinetics of the liquid motion in respect with the gravity direction. Then, to validate the results in this case, it is necessary to fit the kinetics of the liquid motion along the porous sample by comparing with the mass gain curves from FEM and/or experimental data at each separate zone regarding the gravity direction of the porous sample.

## 5. References

[1] L. Bergamaschi and M. Putti, 'Mixed finite elements and Newton-type linearizations for the solution of Richards' equation', *Int. J. Numer. Methods Eng.*, vol. 45, no. 8, pp. 1025–1046, Jul. 1999.
[2] R. Kodešová, J. Šimůnek, A. Nikodem, and V. Jirků, 'Estimation of the Dual-Permeability Model Parameters using Tension Disk Infiltrometer and Guelph Permeameter', *Vadose Zone J.*, vol. 9, no. 2, p. 213, 2010.
[3] A. K. Nguyen, E. Blond, T. Sayet, A. Batakis, E. de Bilbao, and M. D. Duong, 'Self-organized gradient percolation method for numerical simulation of impregnation in porous media', *Comput. Methods Appl. Mech. Eng.*, vol. 344, pp. 711–733, Feb. 2019.
[4] *Computer Vision and Applications*. Elsevier, 2000.
[5] P. A. Vermeer and A. Verruijt, 'An accuracy condition for consolidation by finite elements', *Int. J. Numer. Anal. Methods Geomech.*, vol. 5, no. 1, pp. 1–14, Jan. 1981.
[6] G. Grimmett, *Percolation*, vol. 321. Berlin, Heidelberg: Springer Berlin Heidelberg, 1999.
[7] P. Nolin, 'Critical exponents of planar gradient percolation', *Ann. Probab.*, vol. 36, no. 5, pp. 1748–1776, Sep. 2008.